# Hydrodynamics shapes annularity in coral reefs via scale-free growth processes

Eva Llabrés*[1,2], Àlex Giménez-Romero[3,1], Tomàs Sintes[1], Carlos M. Duarte[4]

*Corresponding author. Email: evallabres@ifisc.uib-csic.es

[1]Institute for Cross-Disciplinary Physics and Complex Systems, IFISC (CSIC-UIB), Universitat de les Illes Balears, 07122 Palma de Mallorca, Spain

[2]Hawai'i Institute of Marine Biology, University of Hawai'i at Manoa, Kaneohe, HI 96744, USA

[3]CEAB (CSIC), Department of Ecology and Complexity, Theoretical and Computational Ecology Group, Blanes, Catalonia, Spain

[4]Marine Science Program, Biological and Environmental Science and Engineering Division, King Abdullah University of Science and Technology (KAUST), Thuwal, Saudi Arabia

## Abstract

Atolls are traditionally explained as the result of coral reefs accreting around volcanic islands followed by gradual subsidence, yielding a hollow, ring-shaped rim that can extend for kilometres. However, satellite imagery shows that similar annular outlines also appear in much smaller patch reefs, where atoll-forming geological pathways do not apply. In some systems, small annular patches occur within the lagoons of larger atolls, producing nested ring-like patterns. The recurrence of annularity across such contrasting contexts and scales suggests that shared, self-organising processes may also contribute to shaping these reefs. Here, we test whether interactions between reef growth and marine currents can generate annular forms and explain their cross-scale geometric regularities. We develop a numerical model in which coral growth follows simple process-based rules, with local colonisation and mortality depending on resource supply and hydrodynamic stress, and water flow resolved using fluid dynamics. Simulations show that this coupling robustly produces ring-like patch reefs and atoll-like configurations across spatial scales, consistent with observed morphologies. Beyond qualitative agreement, the emergent reefs reproduce key geometric signatures reported in global datasets, including scaling laws and fractal dimensions. Together, these results identify coral-current interactions as a plausible pathway to annular reef formation and a mechanistic explanation for scale-free reef geometry.

## Keywords

Coral reefs, self-organization, pattern formation, hydrodynamic transport, scale-free geometry, agent-based modelling.

# 1. Introduction

Coral reefs owe much of their ecological significance to the carbonate frameworks they build, which create multilayered habitats that sustain exceptional biodiversity and productivity (Graham et al., 2014) while also protecting shoreline (Ferrario et al. 2014). These reef frameworks accrete over millennial timescales, hence precluding direct observation of their formative processes, which have traditionally been interpreted through a geological lens, with emphasis on antecedent templates and pre-existing substrate (Schlager, 2005). This geological perspective is especially prominent in attempts to explain atolls, annular reef rims that encircle a central lagoon. Their formation is classically explained by Darwin's subsidence theory, in which corals first grow around a volcanic island to form fringing or barrier reefs, and as the island slowly sinks, reef growth keeps pace with sea level, ultimately leaving a ring-shaped atoll (Darwin, 1889).

Darwin's subsidence theory remains a cornerstone of reef geology, offering a robust explanation for the formation of many atolls (Ladd et al., 1953, 1970). However, the hypothesis does not apply to similar structures in non-volcanic settings, and a growing body of work shows that annular reef rims are not always straightforward outcomes of subsidence alone, and their origin remains actively debated (Droxler & Jorry, 2021). One prominent example is the Maldivian archipelago, where atolls are built atop a broad platform rather than a single isolated seamount, so their annularity cannot be attributed to a single subsiding volcano (Betzler et al., 2009). Notably, Maldivian reefs conform to a nested structure where atoll rims enclose wide lagoons that themselves host circular, sand-filled patch reefs, producing nested ring-like patterns across scales (Fig. 1b). Comparable hollow morphologies recur in very different settings: in Kāneʻohe Bay (Hawaiʻi), many ~200 m patch reefs are roughly circular, with outer zones of higher coral cover surrounding comparatively depleted interiors (Fig 1a), and within the lagoon of Heron Reef (Great Barrier Reef), numerous ~20-m patch reefs likewise exhibit annular forms (Fig. 1c). At still smaller scales, individual colonies can form annular 'microatoll' structures on the order of a few metres, providing a single-organism analogue of the same hollow-ring geometry (Stoddart et al. 1979). Taken together, these examples show that annularity can recur without a consistent geomorphic template, pointing to additional drivers for its emergence (e.g., Jokiel, 1991; Schlager & Purkis, 2015). Moreover, the presence of annular forms across orders of magnitude in size suggests a potentially scale-invariant organisation in coral reefs, consistent with recent global satellite-based analyses reporting shared universal geometric structure across spatial scales and coral provinces (Giménez-Romero et al., 2025). Yet, to our knowledge, annular reef morphologies have rarely been analysed as a connected phenomenon or framed within a unifying mechanism.

Recurrent spatial patterns of this kind are often interpreted in ecology as signatures of self-organization, emerging from interactions between organisms and their environment rather than from imposed templates (Rietkerk & van de Koppel, 2008). In many ecosystems, transport processes such as water flow play a central role in mediating these feedbacks, redistributing resources and stress in ways that can optimize growth and survival. Well-known examples include dryland vegetation patterns arising from processes that concentrate water and nutrients (Rietkerk et al., 2002), and banded structures in intertidal mussel beds shaped by flow-driven algal delivery (van de Koppel et al., 2005, 2008). Water currents may play an analogous role in coral reefs, since hydrodynamic flows regulate the delivery of key resources, such as dissolved inorganic carbon, oxygen, or particulate food, while imposing mechanical loads for colonies exposed to high current (Comeau et al., 2014; Houlbrèque & Ferrier-Pagès, 2009; Mass et al., 2010; Madin et al, 2014). This creates a trade-off in which water motion can enhance resource supply yet simultaneously constrain reef development through physical disturbance (Grigg, 1998). These coupled effects make coral reefs a plausible candidate for flow-mediated self-organization, in which spatial differences in exposure and resource delivery could translate into systematic structure at the scale of reefscapes (Mistr & Bercovici, 2003; Rietkerk & van de Koppel, 2008).

If coral-flow interactions can organise reefscapes, then understanding when and how they generate particular patterns becomes essential, both for understanding the fundamentals of reef morphogenesis and for anticipating responses to environmental change and guiding conservation strategies. Because these spatial feedbacks cannot be observed along human life spans and are difficult to infer directly from field data, mathematical and numerical models have become key tools for studying self-organization in ecological systems. Existing coral-reef models have provided valuable insights—for example, how nonlinear demographic processes can generate multiple reef morphologies (Álvarez-Alegría et al., 2025), how coupling between growth and currents can impose characteristic orientations or spacings among reefs (Mistr & Bercovici, 2003), or how flow-mediated interactions can produce reticulated structures within reef interiors (Xi et al., 2025). Nevertheless, reef-scale modelling efforts remain relatively scarce, and few frameworks have explicitly addressed the origin of recurrent annular morphologies, their relationship with hydrodynamic forcing, or their connection to the scale-free geometric properties observed across global reefscapes.

Here we test whether the interaction between water currents and reef growth is a plausible mechanism underlying the emergence of annular reef morphologies documented across coral provinces and spatial scales. We develop a parsimonious agent-based model of coral growth coupled to a physics-based description of water flow and resource transport. The model is defined by a parsimonious set of local rules: reefs expand clonally with probabilities weighted by resource availability, while mortality arises from resource limitation and hydrodynamic stress. These demographic dynamics alter reef geometry, which in turn reshapes the surrounding flow field and redistributes resources. As a result, the coupled dynamics naturally produce ring-like morphologies resembling those observed across the world's reef systems. Beyond qualitative agreement, the model reproduces key geometric signatures reported in global reef data—including area-perimeter scaling and fractal measures—closely matching the empirically-resolved traits reported by Giménez-Romero et al. (2025). Together, our results provide a mechanistic interpretation for these regularities and support coral-flow coupling as a candidate driver of annular reef organisation, helping guide coral conservation and restoration efforts.

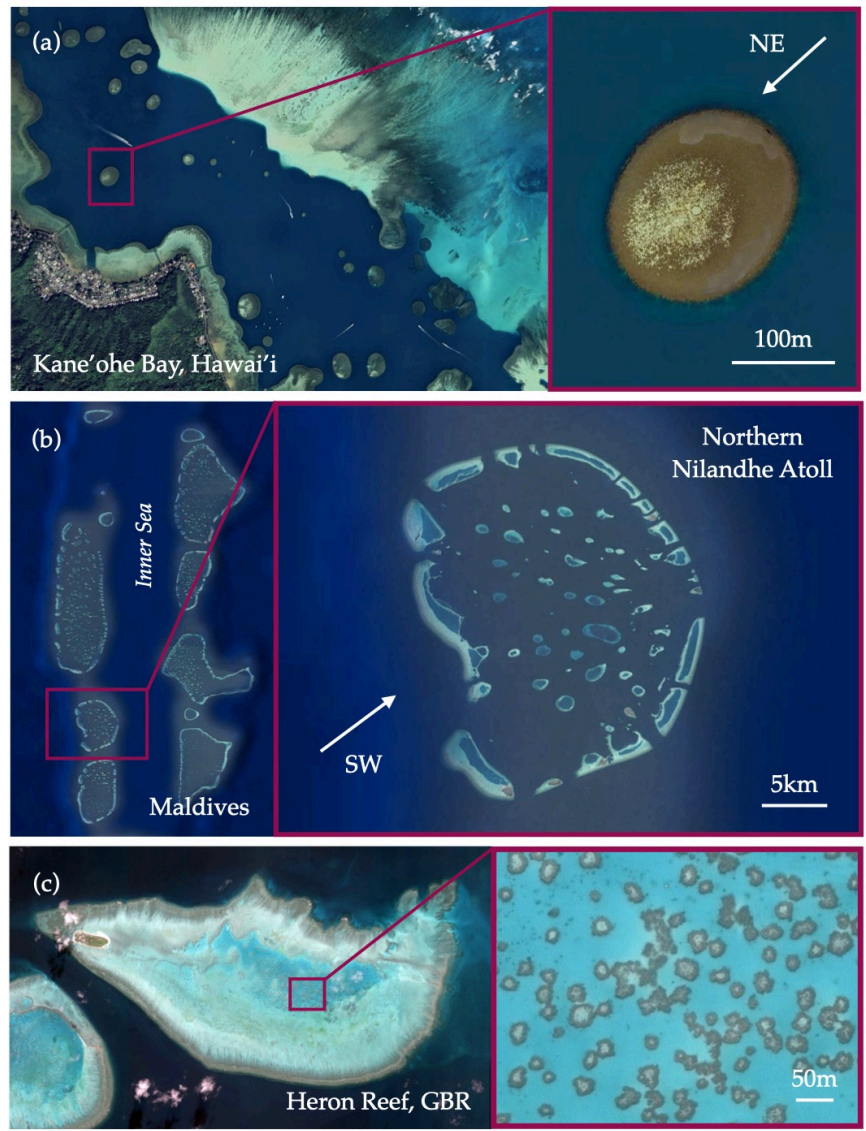

**Figure 1. Examples of annular reef morphologies across regions and spatial scales. (a)** Patch reefs in Kāneʻohe Bay, Hawaiʻi. **(b)** Atolls in the Maldives. **(c)** Lagoonal patch reefs within Heron Reef, Great Barrier Reef. Magenta boxes and connecting lines indicate areas that are magnified in the right-hand panels to provide closer views of representative reef structures. See Supplementary Fig. S1 for additional close-up examples of Kāneʻohe Bay patch reefs. White arrows indicate the dominant current forcing in each region: (a) patch reefs located near the centre of Kāneʻohe Bay are primarily influenced by north-east (NE) trade-winds (Jokiel, 1991); (b) Northern Nilandhe Atoll is predominantly affected by south-west (SW) monsoonal currents (Naseer & Hatcher, 2001). Approximate scale bars are shown as horizontal white lines. Satellite imagery was obtained from Google Earth Pro (acquisition years 2018-2024).

## 2. Materials and Methods

Here we describe the numerical framework used to test whether coral-flow interactions alone can generate annular reef morphologies and scale-free growth. We first present the model (Sec. 2.1), then define the metrics used to quantify the geometry of the simulated reefs (Sec. 2.2).

### 2.1. Numerical coral-flow model

The model is organised as a closed coral-flow feedback loop, schematically summarised in Fig. 2. At any given time, the reef exists as a discrete spatial configuration on a two-dimensional lattice (Fig. 2A). This configuration defines the geometry through which water flows and resources are transported. For a fixed reef state, we compute the surrounding hydrodynamic and resource fields by solving a two-dimensional flow and transport problem (Fig. 2B), treating coral as solid obstacles that locally impede flow and remove resources. These continuous fields are then used to update the lattice through stochastic demographic rules acting at the scale of individual grid cells (Fig. 2C). Changes in reef occupancy alter the flow in the next cycle, closing the feedback between coral demography and hydrodynamics. Repeated iteration of this loop gives rise to emergent reef morphologies across spatial scales. Below, we describe the implementation of this framework in detail.

#### 2.1.1. Biologically-based coral growth rules

Reef growth and loss are implemented through stochastic update rules applied to a two-dimensional lattice at discrete time intervals of duration $\Delta t$ (schematically illustrated in Fig. 2C). At each update step, lattice occupancy is modified by local demographic events that depend on the surrounding physical environment. These events are mediated by two spatially varying quantities (or fields) defined at each lattice site $\vec{x}$: a scalar resource concentration $R(\vec{x})$, representing an aggregate measure of resource availability (e.g., dissolved inorganic carbon, oxygen, and particulate food), and a local flow velocity field $\vec{u}(\vec{x})$, which captures hydrodynamic exposure. Both $R$ and $\vec{u}$ are computed for each reef configuration by the hydrodynamic solver described in Section 2.1.2. Event probabilities are defined as local functions of these fields, so that demographic updates depend explicitly on the values of $R(\vec{x})$ and $\vec{u}(\vec{x})$ at each site, following the general approach of seminal works, such as Niemeyer et al., 1984. At each update step, lattice occupancy is modified through the following stochastic processes:

1. **Clonal expansion weighted by resource availability:** With probability $\gamma$, one candidate cell is chosen from the set of empty lattice sites that are face-adjacent to existing coral (empty squares in Fig. 2). Selection is weighted by local resource availability so that the probability of picking site at $\vec{x}$:

$$p_{\text{clone}}(\vec{x}) = \gamma \frac{R(\vec{x})^{\xi}}{\sum_{\vec{x}' \in \square} R(\vec{x}')^{\xi}} \qquad (1)$$

   where $\xi$ regulates the degree to which resource gradients bias selection among eligible sites, and the denominator normalises the distribution over all eligible growth sites ($\square$). Once selected, the chosen site becomes occupied.

2. **Mortality due to erosion:** With probability $\omega_{\text{ero}}$, a cell is selected from the currently occupied coral cells (black squares in Fig. 2). The likelihood of removing the colony at position $\vec{x}$ scales with the local hydrodynamic stress,

$$p_{\text{ero}}(\vec{x}) = \omega_{\text{ero}} \frac{|\vec{u}(\vec{x})|^{\xi}}{\sum\limits_{\vec{x}'\in\blacksquare} |\vec{u}(\vec{x}')|^{\xi}} \tag{2}$$

The denominator normalises the distribution over all occupied sites (■), so that colonies exposed to stronger currents are more likely to be lost.

3. **Mortality due to resource scarcity**: With probability $\omega_{\text{res}}$, one occupied cell is removed, with selection weighted towards locations with low resource availability. We first compute a resource-weighted distribution over occupied sites:

$$p(\vec{x}) = \frac{R(\vec{x})^{\xi}}{\sum\limits_{\vec{x}'\in\blacksquare} R(\vec{x}')^{\xi}} \tag{3}$$

and then invert and renormalise it to obtain the mortality probability,

$$p_{\text{mort}}(\vec{x}) = \omega_{\text{res}} \frac{1 - p(\vec{x})}{\sum\limits_{\vec{x}'\in\blacksquare} (1 - p(\vec{x}'))} \tag{4}$$

This assigns higher removal probabilities to occupied sites in resource-poor regions.

In all three processes, the exponent $\xi$ controls how strongly demographic events respond to spatial heterogeneity in the environmental fields: for $\xi \to 0$, selection is nearly uniform among candidate sites, whereas larger $\xi$ increasingly concentrates events on the most favourable locations. Throughout this work we fix $\xi = 0.5$, representing an intermediate, moderately selective regime that preserves sensitivity to environmental gradients without producing overly localised dynamics.

The parameters $\omega_{\text{ero}}$, and $\tilde{\omega}_{\text{res}}$ therefore weight the relative importance of clonal growth, erosion-driven loss, and resource-driven mortality, respectively, and jointly control the emergent reef morphologies. For simplicity, we will express mortality rates relative to the clonal expansion probability and define the dimensionless ratios:

$$\tilde{\omega}_{\text{res}} = \omega_{\text{res}}/\gamma \qquad \tilde{\omega}_{\text{ero}} = \omega_{\text{ero}}/\gamma \tag{5}$$

which measure resource- and erosion-driven loss as fractions of the clonal expansion. In this formulation, $\gamma$ primarily sets the pace of the stochastic updates. Because our focus is on the emergent morphologies rather than on absolute growth rates, we do not map $\Delta t$ (and thus $\gamma$) to empirical time units here, and leave such calibration for future work. Unless stated otherwise, simulations were run for 300 update sweeps, each consisting of a full pass of the stochastic demographic rules over the set of growth-eligible sites.

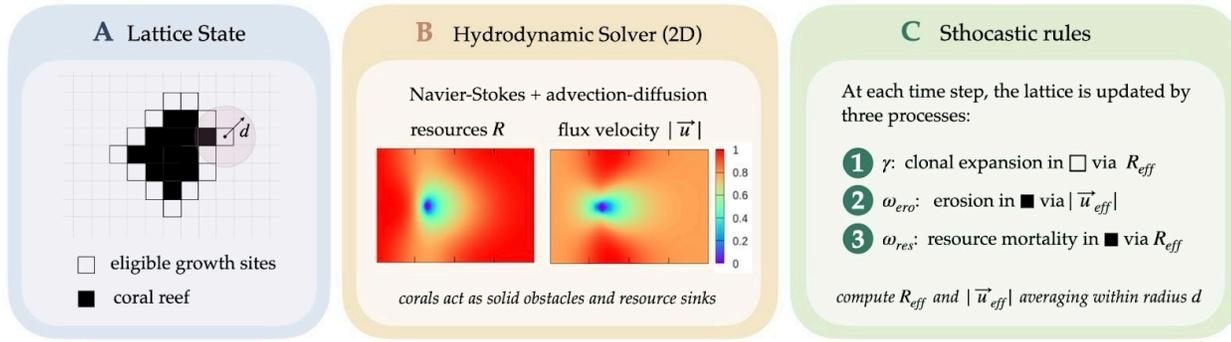

**Figure 2. Schematic overview of the coral-flow feedback model. (A)** The reef is represented on a two-dimensional lattice, where black cells denote occupied coral and white cells indicate eligible growth sites. A neighbourhood of radius $d$ (pink circle) defines the scale over which effective exposure is evaluated. **(B)** For a given reef configuration, the two-dimensional water flow and resource fields are computed by solving the incompressible Navier-Stokes equations coupled to an advection-diffusion equation using a lattice-Boltzmann method. Coral cells act as solid obstacles and local resource sinks, generating spatial gradients in resource concentration $R$ and flow speed $\vec{u}$, as illustrated by the example fields shown in the panel. **(C)** Reef occupancy is updated by three stochastic processes: (1) clonal expansion on empty growth sites with probability $\gamma$; (2) erosion of occupied coral with probability $\tilde{\omega}_{\text{ero}}$; and (3) mortality due to resource limitation with probability $\tilde{\omega}_{\text{res}}$. Event selection is weighted by effective fields $R_{\text{eff}}$ and $\vec{u}_{\text{eff}}$, computed by averaging $R$ and $\vec{u}$ over neighbouring cells at distance $< d$, as illustrated in (A). The updated lattice then defines the reef geometry for the next hydrodynamic solve, and the cycle repeats.

### 2.1.2. Hydrodynamic and resource-transport model

To resolve fluid flow and nutrient transport fields that mediate these local growth and mortality processes, we employ a physics-based framework. The velocity field $\vec{u}(\vec{x})$ and resource concentration $R(\vec{x})$ are obtained by numerically integrating the incompressible Navier-Stokes equations coupled to an advection-diffusion, using a two-dimensional approximation:

$$\nabla \cdot \vec{u} = 0 \qquad \left(\frac{\partial \vec{u}}{\partial t} + (\vec{u} \cdot \nabla)\vec{u}\right) = -\frac{1}{p}\nabla p + \nu \nabla^2 \vec{u} \qquad \frac{\partial R}{\partial t} + (\vec{u} \cdot \nabla)R = D\nabla^2 R \qquad (6)$$

where $p$ is the pressure, $\rho$ the density, and $\nu$ and $D$ are the kinematic viscosity and scalar diffusivity, respectively.

Throughout this work, we restrict attention to a physically steady flow regime in which scalar transport is primarily advection-driven, and we do not attempt to represent turbulence, wave-driven oscillatory flows, or other short-timescale variability. For each fixed reef configuration, we solve the coupled hydrodynamic and transport equations using the lattice Boltzmann method (LBM) with a Bhatnagar-Gross-Krook (BGK) collision model (Succi, 2001; Krüger et al., 2017). Beyond being computationally efficient, the LBM-BGK formulation supports consistent rescaling across spatial scales. To test reef growth across scales while preserving dynamically equivalent flow conditions, we simulate hydrodynamics under three alternative spatial scalings (Cases I-III; Table S1). More details of the hydrodynamic regime and rescaling procedure are provided in Supplementary Materials A.

We implement the hydrodynamic solver in a rectangular domain with standard boundary conditions implemented in the LBM-BGK scheme (Zou & He, 1997). Flow enters from the left as a uniform inflow carrying a fixed resource concentration and exits through an open outflow boundary on the right, while the upper and lower boundaries are periodic. Occupied lattice sites are treated as impermeable obstacles, imposing a no-slip condition at their surface ($\vec{u} = 0$), and as perfect sinks for the transported scalar $R = 0$. Because natural reefs are porous at scales below our grid resolution, we approximate partial exchange by defining *effective* fields used in the biological update rules (Fig. 2C). For each coral site at position $\vec{x}$, we compute neighbourhood-averaged resource availability and flow exposure within a radius $d$, as illustrated schematically in Fig. 2A,

$$R_{\text{eff}}(\vec{x}) = \frac{1}{N_d} \sum_{|\vec{x}-\vec{x}'|<d} R(\vec{x}') \qquad |\vec{u}_{\text{eff}}(\vec{x})| = \frac{1}{N_d} \sum_{|\vec{x}-\vec{x}'|<d} |\vec{u}(\vec{x}')| \qquad (7)$$

where $N_d$ is the number of lattice sites within the averaging neighbourhood. The parameter $d$ sets the spatial scale over which porosity is represented: larger $d$ smooths the sharp gradients created by impermeable obstacles, while smaller $d$ approaches the impermeable limit. For convenience, we also introduce the dimensionless permeability distance, where $\tilde{d} = d/\Delta x$, where $\Delta x$ is the lattice cell size. To ensure consistency across spatial scalings, $d$ is chosen separately for each study case (Cases I-III, Table S1). These effective fields are used only to evaluate growth and mortality probabilities, and do not modify the hydrodynamic or transport equations themselves.

## 2.2. Spatial properties of coral reefs

In this work, we quantitatively compare the output of our model with observations using established geometric metrics. Giménez-Romero et al. (2025) recently analysed reef morphology at the global scale and reported universal geometric relationships that hold across coral-reef provinces. Their analysis uses reef outlines extracted from the Allen Coral Atlas (2022), a global high-resolution dataset derived from satellite imagery using machine-learning-assisted habitat classification. Here we test whether reefs generated by our coral-flow model reproduce the same universal regularities. Specifically, we quantify the scaling relationship between reef area $A$ and perimeter $P$ of simulated and observed reefs, and assess whether it follows a power law of the form

$$A \propto P^\alpha \qquad (8)$$

where $\alpha$ is the scaling exponent. For smooth, compact shapes $\alpha = 2$ following the area relation of a circle, while lower values indicate increasing boundary complexity. We also quantify fractality via the area box-counting fractal $D_A$ (Mandelbrot, 1983). For a given box size $\varepsilon$, we overlay the lattice with a square grid and count $N(\varepsilon)$, the minimum number of boxes of size $\varepsilon$ needed to cover the set of occupied cells. The area box-counting fractal dimension is then defined as

$$D_A = \lim_{\varepsilon \to 0} \frac{\log N(\varepsilon)}{\log(1/\varepsilon)} \qquad (9)$$

In practice, the limit is estimated from the slope of the linear fit to $\log N(\varepsilon)$ versus $\log(1/\varepsilon)$ over a range of scales. A value $D_A = 2$ corresponds to a fully filled, smooth shape, with smaller values indicating irregular and less space-filling structures (Mandelbrot 1983). Together, the fractal dimension $D_A$ and the area-perimeter exponent $\alpha$ provide complementary descriptors of reef morphology that enable direct, quantitative comparison between simulated reefs and globally observed reef geometries.

# 3. Results

We first qualitatively examine the morphologies produced by the model under a representative set of parameters chosen to illustrate distinct reef-growth scenarios (Section 3.1). We then systematically vary the mortality parameters $\tilde{\omega}_{res}$ and $\tilde{\omega}_{ero}$ to quantify their effects on reef geometry and to compare model outputs with spatial properties observed in global reef data (Section 3.2).

## 3.1. Simulated reef morphologies across spatial scales

Our main results are summarized in Fig. 3, where we show that the model reproduces reef morphologies at different scales resembling those observed in satellite imagery (see Fig. 1). Here we note this qualitative correspondence, while its geological and ecological interpretation are developed in the Discussion (Sec. 4.1).

We begin with a single-seed simulation that gives rise to a solitary patch reef (Fig. 3a-c). Growth initially proceeds radially, forming an approximately compact circular structure. However, as the reef expands, interior regions experience resource depletion due to limited penetration of resources (modulated by the permeability distance $d$). Outer-edge colonies, which are better exposed to incoming flow, proliferate clonally, leading to the emergence of an oval, ring-shaped reef (Fig. 3b-c). Under unidirectional flow, the ring becomes slightly asymmetric, with a thickened flow-facing rim and a thinner back side. This morphology closely resembles the patch reefs shown in Fig. 1a.

Next, we explore simulations initiated from multiple seeds (Fig. 3d-f). Early on, reefs grow independently and develop ring-like morphologies similar to the single-seed case (Fig. 3d). As they expand and begin to interact, a clear upstream-downstream contrast emerges, and reefs on the flow-facing front receive the highest nutrient flux and outgrow those on the downstream back (Fig. 3e). At later stages, the pattern self-organizes into an approximately circular ring-like configuration, with the most elongated reefs located at the flow-facing side (Fig. 3f). Superimposed on this system-scale asymmetry, individual reefs along the ring also develop a second level of asymmetry, as their outer-facing crests thicken relative to their lagoon-facing sides. Reefs throughout this system remain predominantly hollow at all stages. As a result, the emergent atoll-like structure exhibits a nested circular organization: the outer ring is itself composed of hollow reefs, enclosing a lagoon that contains smaller, similarly hollow patches. Taken together, these morphologies resemble atoll-like patterns such as those observed in the Maldives (Fig. 1c).

Finally, we examine scenarios with many initial seeds and reduced erosion (Fig. 3g-i). In this regime, reduced erosion serves as a proxy for settings where disturbances are weak, such as sheltered and shallow lagoonal environments. The resulting structures are small, nearly circular patch reefs (≈20 m in diameter), often hollow in the center, resembling the lagoonal formations observed in Fig. 1b. Unlike the interacting reefs in Fig. 3d-f, growth proceeds mostly independently, and competitive asymmetries do not emerge, at least at the spatial extent considered.

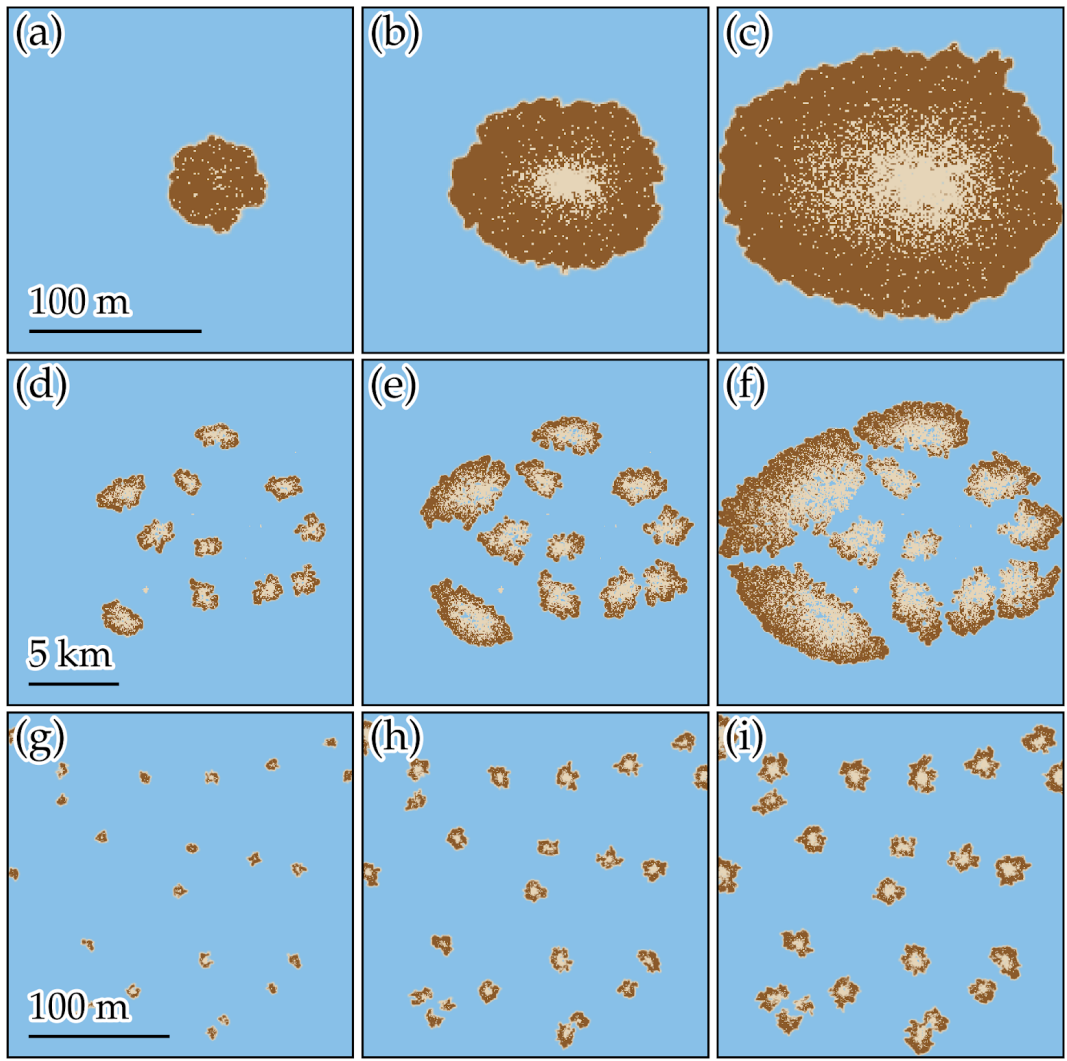

**Figure 3. Simulated reef morphologies generated by the coral growth model coupled to a steady left-to-right incoming current**. **(a-c)** Single-seed growth gives rise to a hollow patch reef. **(d-f)** Multiple seeds expand and interact, producing atoll-like chains with upstream-downstream asymmetries. **(g-i)** Numerous seeds with weak erosion yield small, nearly circular lagoonal reefs that remain isolated. Reef occupancy is shown as dark brown for currently occupied cells (live coral), light brown for previously occupied but now unoccupied cells (dead coral, rubble or sand), and blue for cells that were never occupied (seafloor). Scale bars indicate physical distances. Columns show temporal progression from left to right. Rows correspond to the three hydrodynamic cases (I-III), which vary the lattice spacing while preserving the same flow-transport regime (Table S1).

### 3.2. Morphological regimes and geometric scaling

In the previous subsection, we showed that annular reef morphologies emerge for specific combinations of resource-driven mortality ($\tilde{\omega}_{\text{res}}$) and erosion ($\tilde{\omega}_{\text{ero}}$). We now quantify how reef morphology responds systematically to changes in these a-dimensional parameters (Fig. 4a). Low $\tilde{\omega}_{\text{res}}$ yields compact, filled reefs with little interior mortality. As $\tilde{\omega}_{\text{res}}$ increases, interior resource limitation intensifies, leading to progressive central die-off and the emergence of hollow, ring-like structures, as already illustrated in Fig. 3a-c. By contrast, higher $\tilde{\omega}_{\text{ero}}$ smooths reef boundaries, as exposed edge colonies are more vulnerable to erosion. Beyond a certain threshold, reefs become elongated along the flow direction, consistent with elevated mortality along the lateral flanks where flow velocities are highest (Fig. 2B). Crucially, annular morphologies appear only within a subset of parameter space, where interior depletion is strong enough to hollow reefs, and erosion is intermediate—sufficient to smooth the rim, but not so strong that it suppresses lateral growth and drives streamwise elongation.

Notably, area-perimeter power laws emerge robustly across the entire parameter space, regardless of whether the final reef morphology is compact or hollow (Fig. 4b). At small sizes, all simulations follow a common scaling regime, with an exponent close to the compact limit ($\approx 1.8$), largely independent of mortality parameters. This regime corresponds to the early growth phase, when reefs expand as space-filling patches, matching the morphologies observed in Fig. 3a. Beyond a characteristic perimeter, reefs transition to a second scaling regime in which the exponent $\alpha$ depends sensitively on the balance between erosion and resource-driven mortality. This transition point is set by the permeability length $d$: once reefs become large compared with this penetration distance, growth is effectively confined to a rim (Fig. 3b-c), and the scaling changes. The background heatmap in Fig. 4a shows that the late-regime exponent $\alpha$ is smallest when resource limitation is the primary driver of colony loss, and largest when erosion dominates. This trend is consistent with the morphology map and has a simple geometric interpretation: when reefs are compact and space-filling (low $\tilde{\omega}_{res}$, high $\tilde{\omega}_{ero}$), their area grows nearly as the square of the perimeter, so $\alpha \approx 2$, approaching the compact circle-like limit. By contrast, hollow or highly indented reefs (high $\tilde{\omega}_{res}$, low $\tilde{\omega}_{ero}$) have disproportionately large perimeter for their enclosed area, shifting the scaling toward smaller exponents, for which the limits are filamentary, ribbon-like geometries, with $\alpha \approx 1$.

The two-phase scaling observed in our simulations closely parallels that reported for natural reefs worldwide (Fig. 4c). Closer inspection of global inventory compiled by Giménez-Romero et al. (2025), shows that smaller reefs follow a power-law with exponent $\approx 1.62$, broadly consistent with the first-regime scaling for compact shapes found in our simulations ($\approx 1.83$; Fig. 4b). In the second regime, simulated exponents fall within the empirical range only for a region of the parameter space. In particular, the parameter combinations near $\tilde{\omega}_{res} \approx 0.5$ and $\tilde{\omega}_{ero} \approx 0.2$ yield a late-regime exponent $\alpha \approx 1.18$ (Fig. 4b), in close agreement with the global estimate ($\approx 1.26$; Fig. 4c), while also generating annular morphologies characteristic of real reefs (Fig. 1). This agreement provides an internal consistency check, indicating that the most plausible parameter values of $\tilde{\omega}_{res}$ and $\tilde{\omega}_{ero}$ are those that jointly capture both the qualitative emergence of annular structures and the quantitative area-perimeter scaling. Moreover, our separate fits for small and large perimeters, an analysis not made in the original work by Giménez-Romero et al. (2025), formalises their observation that reef shapes tend to shift from compact shapes at small sizes to increasingly boundary-dominated geometries at larger sizes.

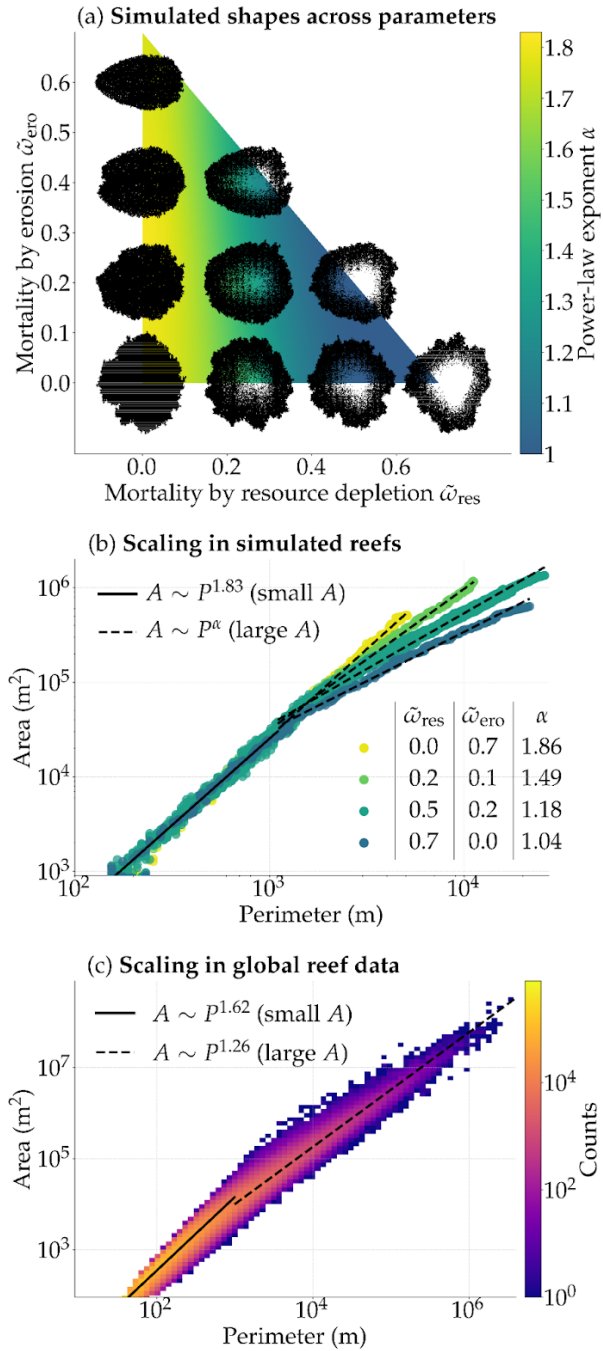

**Figure 4. Sensitivity of reef morphology to mortality parameters**. **(a)** Simulated morphologies across parameter space ($\tilde{\omega}_{\text{res}}, \tilde{\omega}_{\text{ero}}$). Each black silhouette is centred at the parameter values used for that simulation. Flow is imposed from left to right and hydrodynamic setup fixed as Case I in Table S1. Only parameter pairs with $\tilde{\omega}_{\text{ero}} + \tilde{\omega}_{\text{res}} < 0.7$ are shown, since larger sums approach the extinction threshold. **(b)** Area-perimeter scaling for four representative parameter combinations. For each combination, $(A, P)$ pairs were sampled over the growth trajectory of the shapes and aggregated across stochastic realisations. All cases exhibit two scaling regimes: an early compact-growth regime at small sizes ($\approx 1.83$, solid line) followed by a late regime with parameter-dependent exponent $\alpha$ (dashed lines). Colours correspond to the parameter sets listed in the inset table. The background heatmap in (a) reports the late-regime exponent $\alpha$ across the full parameter space. **(c)** Empiric area-perimeter relations from global reef data sets (Giménez-Romero et al. 2025) showing also two-phase scaling.

We extended the analysis by considering simulations initiated from multiple seeds, allowing us to examine how different combinations of erosion ($\tilde{\omega}_{\text{ero}}$) and results-driven mortality ($\tilde{\omega}_{\text{res}}$) shape the collective dynamics of reef islands (Fig. 5). As in Fig. 4, increasing $\tilde{\omega}_{\text{res}}$ promotes interior die-off and the formation of hollow structures, whereas reducing $\tilde{\omega}_{\text{ero}}$ enhances boundary complexity and produces more irregular contours. In the multi-seed setting, however, the effects of boundary roughness are magnified: when erosion is weak, we observe the emergence of branched, flow-oriented structures reminiscent of dendritic growth (Fig. 5; zoomed in detail in Fig. S3). The amplification of boundary complexity in Fig. 5 relative to Fig. 4 reflects the different parameter set consider (see Table SM1)—specifically the smaller adimensional permeability distance $\tilde{d}$ sharpens resource absorption at reef edges and favors branches that extend in the direction of the current. This mechanism is examined in more detail in the Discussion (Sec 4.3). Despite this, as in the single seed case, parameter combinations around the values $\tilde{\omega}_{\text{res}} \approx 0.5$ and $\tilde{\omega}_{\text{ero}} \approx 0.2$ still produce hollow annular structures arranged in approximately circular configurations, closely resembling natural atoll chains.

To quantify these multi-seed patterns, we evaluated the fractal dimension of the occupied area, $D_A$, as an indicator of how efficiently simulated reefs fill space. Across three representative mortality scenarios, the $D_A$ curves show similar overall trends (Fig. 5b). Because these curves represent binned averages, differences between scenarios are subtle; clearer contrasts emerge in the distributions and mean values. When erosion is strong, reefs tend to be fewer, larger, and more compact, producing narrower and shorter histograms and correspondingly higher averages $\langle D_A \rangle$. Despite the wide variety of reef shapes generated, the average fractal dimension remains confined to a narrow range $D_A \approx 1.4 - 1.6$ (heatmap in Fig. 5a). These values are very similar to the global average observed in nature $D_A \approx 1.6$ calculated by Giménez-Romero et al. (2025) and replotted in Fig. 5c, hinting at a possible correspondence between model outcomes and natural reef complexity that we examine further in the Discussion (Sec. 4.2).

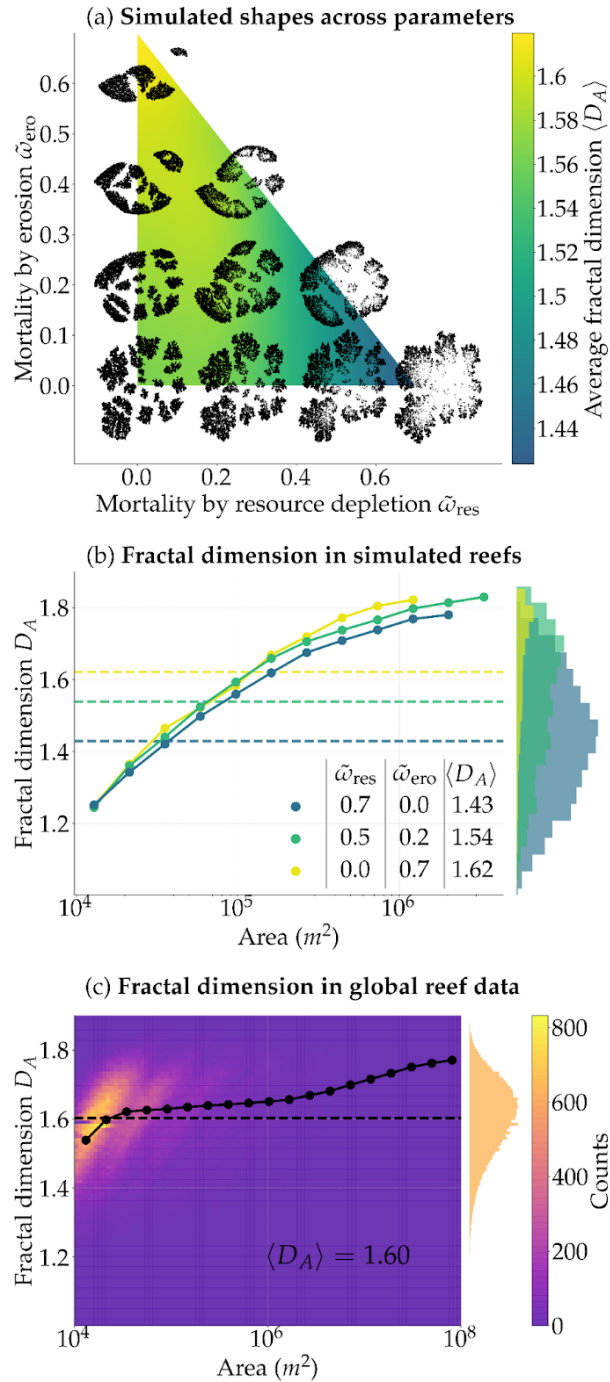

**Figure 5. Fractal dimension $D_A$ of simulated and natural reefs. (a)** Simulated morphologies across parameter space ($\tilde{\omega}_{res}$, $\tilde{\omega}_{ero}$) with background heatmap indicating the mean fractal dimension $\langle D_A \rangle$. Flow is imposed from left to right and hydrodynamic setup fixed as Case II in Table S1. Only parameter pairs with $\tilde{\omega}_{ero} + \tilde{\omega}_{res} < 0.7$ are shown, since larger sums approach the extinction threshold. A zoomed detail highlighting the dendritic morphologies that emerge at low erosion is provided in Supplementary Fig. S3. **(b)** Fractal dimension $D_A$ as a function of reef area for three representative parameter sets. Solid curves show binned averages of $D_A$ versus area, side histograms summarise the distribution of its values across stochastic realizations, and horizontal dashed lines indicate the mean $\langle D_A \rangle$. The inset table links each colour to its corresponding $\tilde{\omega}_{res}$, $\tilde{\omega}_{ero}$, and $\langle D_A \rangle$ value. **(c)** Empirical benchmark from Giménez-Romero et al. (2025), showing the relationship between $D_A$ and area in the global reef dataset; the orange histogram indicates the distribution of $D_A$ across all mapped reefs, and the black dashed line indicates the total average $\langle D_A \rangle$.

## 4. Discussion

Our model produces annular reef morphologies at multiple scales (Figs. 1, 3), as an emergent outcome of coral growth forced by hydrodynamics. These results point to a recurring organisational pattern, in which the living framework is preferentially maintained along exposed margins while interior regions progressively decline. In the following, we develop several implications of these findings. We first place the simulated annular morphologies in the context of classic Darwinian theories for reef formation, and argue that current-mediated coral growth offers an alternative mechanism for the formation of atoll-like structures (Sec 4.1). We then discuss how the same coral-current coupling may help explain the emergence of scale-free geometric properties of reefs (Sec. 4.2), before discussing the limitations of our framework and outline key directions for future work (Sec 4.3).

### 4.1. Non-Darwinian pathways to annular reef formation

As a first implication of our results, we examine how the simulated annular structures relate to long-standing theories of atoll formation. Darwin (1889) proposed that coral reefs grow around a volcanic island to form fringing or barrier reefs. As the island slowly sinks, the corals keep pace with sea level by growing upward, eventually leaving a ring-shaped atoll (Darwin, 1889). While this framework is strongly supported by geological evidence in some settings (e.g., Ladd et al., 1953, 1970), it does not fully account for atoll-like morphologies observed in other regions (Droxler & Jorry, 2021). In such cases, our model offers an alternative, more universal and parsimonious pathway to atoll formation: annular morphologies can emerge from coral-flow interactions alone, without invoking subsidence, sea-level change, or a pre-existing ring-shaped template.

The Maldives provide a clear example where classic Darwinian subsidence alone cannot fully account for atoll morphology (Droxler & Jorry, 2021). The archipelago comprises 26 atoll systems built upon a broad submarine ridge, rather than on individual volcanic cones. Geophysical and stratigraphic evidence indicates that the modern Maldivian atolls emerged through the segmentation and partial drowning of a formerly laterally extensive carbonate platform that developed along the ridge (Betzler et al., 2009). Importantly, this reorganisation did not coincide in time with active volcanism or major tectonic subsidence, but rather with the establishment and strengthening of the Indian Ocean monsoon system (Aubert & Droxler, 1996; Kroon et al., 1991; Zheng et al., 2004, Betzler et al., 2016). On this basis, it has been argued that monsoon-driven current regimes played a central role in reshaping the carbonate platform and organising the present pattern of Maldivian atolls (Betzler et al., 2009). This interpretation aligns naturally with our results: in our simulations, annular and asymmetric reef morphologies emerge from the interaction between persistent currents and coral demography (Fig. 3f), providing a mechanistic pathway to atoll-like structures that does not rely solely on subsidence.

The present-day morphology of Maldivian atolls illustrates how currents can imprint strong signatures on reef growth. The atolls form two north-south chains separated by the Maldives Inner Sea, a shallow, semi-enclosed basin sheltered relative to the open Indian Ocean (Fig. 1b). Across the double chain, ocean-facing margins tend to develop broad, continuous crests, whereas margins oriented toward the Inner Sea are typically narrower and more fragmented (Purdy and Bertram 1993), as exemplified by the zoom in North Nilandhe atoll in Fig. 1b. Naseer & Hatcher (2001) linked this systematic windward-leeward contrast to the seasonally reversing NE and SW monsoons, which modulate wave exposure and current strength across the archipelago. For instance, during the SW monsoon, North Nilandhe's west-facing margin is preferentially exposed while its east-facing side remains sheltered. Atolls on the eastern chain show the complementary pattern, with exposure shifting to the opposite margin. Our results support this interpretation: in our simulations, reefs facing the incoming flow receive higher resource fluxes, sustaining faster growth and forming broad, continuous crests, whereas sheltered reefs receive fewer resources and develop into smaller patches (Fig. 3f). Therefore, the proposed model offers a plausible mechanistic explanation for Maldivian asymmetry, based on current-driven resource transport.

A similar hydrodynamic imprint is evident in the patch reefs of Kāneʻohe Bay, Hawaiʻi, despite its markedly different geomorphic and hydrodynamic setting. The bay is semi-enclosed and wave-sheltered by an outer barrier reef, with circulation primarily driven by persistent north-east trade winds and modulated locally by exchanges through major channels (Jokiel, 1991). Although our model focuses on an idealised, unidirectional flow, it predicts partially depleted interiors and directional rim asymmetries (Fig. 3c), features that are also widespread among Kāneʻohe Bay patch reefs as shown in the supplementary examples (SM Fig. S1). Therefore, the reef highlighted in Fig. 1a should therefore be viewed as representative rather than exceptional. Deviations from idealised annular morphologies—such as irregular hollowing or asymmetry axes that depart from the dominant wind direction—likely reflect additional circulation components not captured by our simplified setup. Taken together with the Maldives case, these observations suggest that persistent currents, even in sheltered environments, can promote interior depletion and imprint directional asymmetries across reef systems.

Our model is also able to reproduce inner-lagoon patterns at a very different scale, such as the small patch reefs observed within the lagoon of Heron Reef on the Great Barrier Reef (Fig. 1c). These patch reefs occur in a shallow, sheltered lagoon and are reported to show no demonstrable link between their spatial organisation and the underlying substrate (Walbran, 1994; Smith et al., 1998; Schlager & Purkis, 2015). In satellite imagery, many patches display hollow, near-circular morphologies with no obvious, systematic directional asymmetry. We recover the same qualitative outcome in our simulations (Fig. 3i): reefs develop annularity while remaining approximately symmetric. In this setting, reefs remain sufficiently small to interact only weakly through flow and resource redistribution, and reduced erosion further limits the amplification of directional growth contrasts. However, other sectors of Heron's lagoon exhibit more ordered, reticulate ridge networks with an apparent directional organisation relative to prevailing flows (Schlager & Purkis 2015, Purkis et al. 2015); which our current formulation does not reproduce. We return to these reticulate patterns in Sec. 4.3, where we discuss extensions of the model that could generate flow-oriented ridge organisation.

### 4.2. Scale-free growth in coral reefs

Beyond specific reef morphologies, our results also shed light on why scale-free geometric relationships emerge so generically in coral reefs. Global analyses reveal that reef area and perimeter obey robust power-law relations across provinces and scales (Fig. 4c), implying the absence of a single characteristic size governing reef shape. Our simulations reproduce this behaviour across diverse morphologies and parameter combinations, and offer a simple tentative mechanistic interpretation.  In our model, scale-free area-perimeter scaling emerges broadly across parameter space, despite substantial variation in reef shape (Fig. 4b).  In the regime where reefs grow into annular structures (Fig. 3b-c), growth proceeds by expanding rings of approximately constant width, determined by the permeability length $d$, while their radius increases over time. In an idealised case, such constant-width annuli naturally generate area-perimeter power laws with exponents close to 1 as scale changes (See SM B.1 for a simple derivation). In our simulations, ring-like reefs exhibit slightly larger exponents (e.g., $\alpha \approx 1.18$; Fig. 4b), consistent with deviations from the ideal case, including irregular hollowing and boundary roughness. Therefore, hydrodynamic transport in our model, sets a characteristic penetration distance $d$ that constrains the thickness of the growing rim, and in turn, produces the observed power-law scaling in reef morphology.

A similar geometric logic may also apply to real reefscapes. From a brief visual inspection of reef outlines worldwide (Allen Coral Atlas, 2022), we noticed that many of the largest reefs appear to cluster into a few recurring planforms—annulus-like atolls, rim-like barrier and fringing reefs, and more convoluted outlines consistent with lagoonal interiors (see Fig. S2 for illustrative examples). Despite their visual differences, these large-scale configurations share an important geometric property: much of the reef area is organised into relatively narrow belts, whose width varies less than their overall extent. Idealising these belts as bands of roughly constant width $w$, simple geometric arguments predict area-perimeter power laws with exponents close to 1 (See SM B for a simple derivation). The correspondence between these empirical patterns and the behaviour of our simulations therefore suggests a plausible geometric origin for the observed power-law scaling, in which hydrodynamically mediated shape selection constrains reef growth into narrow, expanding structures across scales. We are currently testing this hypothesis by developing a formal global classification of reef morphologies and measuring rim widths across belt-type reef types to compare with the effective width $w$ implied by the idealised power-law fit (SM B.3). This effective width $w$ is related to the parameter $d$ in our model, offering a potential route to constrain this transport scale from data and to test the proposed mechanism.

Scale-free patterns have often been interpreted as a hallmark of fractality in other systems, ranging from classic analysis of natural fractals to cities and coastlines (Mandelbrot 1983; Seekell et al. 2013; Mori et al. 2020). In coral reefs, fractal-like organisation has been documented across multiple scales—from colony skeletal roughness to reef outlines (Bradbury & Reichelt 1983; Basillais 1997; Purkis et al. 2007; Zawada & Brock 2009; Zawada et al. 2019; Sous et al. 2020; Giménez-Romero et al. 2025). In our model, this fractal-like behaviour appears in the form of recurrent annular morphologies emerging across scales (Fig. 3), suggesting some degree of self-similarity in reef architecture. This is most evident in the multiple-seed simulations (Fig. 3f), where a single snapshot spans nested spatial scales: small, circular hollow patches persist within larger, elongated (and likewise hollow) reefs that collectively organise into an outer ring, an arrangement observed also in real atolls (Fig. 1b). Moreover, when we lower erosion, the model also generates flow-oriented branching structures (Fig. 4a) that resemble dendritic patterns, which are commonly regarded as fractal (Niemeyer et al. 1984).

To quantitatively assess the fractality of the simulated patterns, we performed a complementary morphometric analysis based on the area fractal dimension $D_A$ (Fig. 5c). Surprisingly, its mean value varies only weakly across parameter space and remains close to $\langle D_A \rangle \approx 1.6$, consistent with the global empirical values reported by Giménez-Romero et al. (2025). This limited variation is unexpected given the markedly different morphologies produced—from compact to strongly dendritic— could suggest that the mean $\langle D_A \rangle$ alone is not a reliable diagnostic of shape type in these settings. Moreover, $D_A$ tends to increase with area (Fig. 5b), even though one might anticipate the opposite: smaller reefs are generally more compact, whereas larger ones can become more complex in shape, a scenario in which one would expect smaller $D_A$. A plausible explanation for both of these counterintuitive results lies in the relatively coarse spatial resolution used in Fig. 5 (cell size ≈ 100 m). Such resolution limits the model's ability to resolve area occupancy, particularly for smaller reefs, and consequently compresses the range of attainable fractal dimensions. Higher-resolution simulations together with a scale-aware analysis of $D_A$ would be required to fully assess these effects, although such computations would be substantially more time-demanding.

### 4.3. Model scope and future directions

Although the model reproduces a broad family of hollow, annular reef structures, several ecological and physical processes remain outside its current scope. These omissions constrain the range of reef architectures that can emerge and limit direct comparison with certain natural settings. Below, we outline key limitations and propose targeted extensions that would broaden the model's applicability and refine its mechanistic foundation.

#### 4.3.1. Extending the model to other reef morphologies

Our model reproduces hollow, annular structures well, but it does not capture several other common reef architectures, including fringing and barrier reefs. These systems typically develop as elongated, shore-parallel reefs developing along coastlines. We likewise do not reproduce the elongated forms reported for many Great Barrier Reef shelf reefs, which have been described as showing systematic alignment with prevailing currents (Mistr & Bercovici 2003, Rietkerk & van de Koppel 2008). Although our simulations can produce belt-like features along atoll perimeters (Fig. 3f), we do not obtain comparable standalone structures whose long axes consistently align with the flow. This limitation arises from two main aspects of our formulation: (i) flow accelerates the most along the flanks of the obstacles, so erosion is strongest precisely where sideways accretion would be required; and (ii) resource supply has high concentrations at the upstream face and depletion along the sides, so growth is promoted into the current and suppressed laterally. These two effects are evident in the numerically solved flow and transport fields shown in Fig. 1B. As a result, at high erosion our simulated reefs tend to elongate along the flow (Figs. 4-5), whereas the reef forms reported on the Great Barrier Reef that are oriented perpendicular to the prevailing current would require growth to be favoured across the current instead. A crucial step to address this issue would be to allow high flux, apart from eroding, to provide positive feedback by clearing fine sediments, reducing turbidity, and thereby improving light conditions and facilitating colony growth (Grigg, 1998). More generally, decoupling the beneficial and damaging effects of flow—so that hydrodynamic exposure can both enhance growth and increase loss—may be necessary to reproduce reef forms that preferentially develop across the current.

Another class of reef patterns that our current formulation does not fully capture are the reticulate, maze-like ridge networks found in many inner-reef lagoons. These were traditionally interpreted as the inheritance of antecedent karst topography, but accumulating evidence suggests that at least a subset can instead arise through ecological pattern formation and biotic self-organisation, much like the reticulate structures observed in mussel beds or arid vegetation systems (Schlager & Purkis 2015). At first glance, the dendritic morphologies that emerge in our simulations at low erosion (Fig. 5, zoomed detail in SM Fig. S3) seemed superficially related to these reticulate reefs. In this regime, branching tips preferentially advance toward the incoming current, reminiscent of dielectric breakdown model (Niemeyer et al. 1984). However, lagoonal dendrites are often reported to organise roughly perpendicular to prevailing currents, although available observations remain sparse (Purkis et al. 2015; Xi et al. 2025). This discrepancy suggests that additional processes favouring lateral, flank-wise propagation— similar to those suggested in the paragraph above—would be needed to also reproduce lagoonal reticulate patterns. If implemented correctly, we would expect reticulate configurations to appear in our simulations as a later growth stage or as a phase transition from the single hollow patch-reef state in Fig. 3g. Indeed, inside Heron Island inner-lagoon one can observe a progression from isolated hollow patches to more reticulate arrangements (Fig. 6 in Schlager & Purkis 2015), reminiscent of similar transitions documented in other self-organised systems such as seagrass meadows (Ruiz-Reynés et al. 2017).

### 4.3.2. Mechanistic foundations of annular reef formation

A central outcome of our model is the emergence of recurrent annular structures, where reef interiors empty out as growth is maintained at the periphery. While we have argued that this mechanism naturally leads to scale-free growth in our simulations (Sec. 4.2), its present formulation is admittedly simplified. In our implementation, reefs act as perfect sinks for advected resources, with permeability encoded through the effective penetration distance $d$ and implemented by summing over lattice cells (Sec. 2.2), but a more physically grounded hydrodynamic formulation would be necessary to relate these dynamics to measurable parameters. Such an approach would involve adding explicit forcing and absorption terms in the hydrodynamics equations (Guo et al., 2002), thereby grounding the model in physical phenomena that can be empirically quantified.

An important dynamical outcome of our 2D framework is that reefs do not converge to a fixed planform: once annularity emerges, they continue to expand outward as a roughly constant-width rim whose radius increases over time. In nature, however, reefs grow in three dimensions, and long-term accretion is often partly redirected into the vertical direction to maintain a favourable depth relative to light, keeping up with sea level changes and subsidence. This vertical accommodation likely limits sustained horizontal expansion in settings such as the Maldives (~100 m relief) and Kāneʻohe Bay (~20 m), so outlines can appear quasi-stationary even while carbonate production remains active. Extending the model to three dimensions by including vertical accretion and accommodation dynamics would allow us to test how vertical growth constrains lateral expansion. Such an extension could be paired with more realistic hydrodynamics that include vertical exchange and wave-driven motions (van de Vijsel et al., 2023; Xi et al., 2025), improving the physical basis of predicted transport and fluxes for direct comparison with natural reef systems.

On the biological basis of the model, mortality in the present formulation is coupled to local resource availability (Sec. 2.1), but corals can also die through other pathways. One important example is age-related deterioration or senescence, which has been documented across multiple coral species (Rinkevich & Loya 1986; Bythell et al. 2018). Incorporating senescence-driven mortality would introduce a complementary, biologically motivated route to interior depletion: colonies in the reef center would eventually die off regardless of resource availability, while peripheral colonies continue to thrive, naturally producing hollow structures.

Incorporating explicit permeability, more realistic hydrodynamics, and senescence-based mortality would thus ground the model more firmly in measurable physical and biological processes. Preliminary (unpublished) simulations indicate that these refinements still generate ring-like geometries emptying dynamics, which will be explored in future work.

## 5. Conclusion

Our goal here was not to reproduce the full mechanistic richness of coral-reef morphogenesis, but to ask whether a parsimonious set of ecological rules coupled to transport can already account for a recurrent pattern: hollow, annular reef outlines that appear across environments and span orders of magnitude in size. By combining (1) growth biased toward higher resource supply, (2) mortality driven by resource limitation, and (3) hydrodynamic stress acting on exposed boundaries, we find that ring-like structures emerge naturally and robustly, without prescribing subsidence or antecedent templates. More broadly, the coupled coral-flow feedback self-organises into geometry that remains consistent across spatial rescalings, providing a simple dynamical route to the scale-free signatures seen in global reef outlines. We therefore view the model as a minimal, process-based baseline for reef annularity against which additional processes—such as vertical accommodation, waves and tides, and sediment-driven light limitation—can be layered and tested.

## References

Allen Coral Atlas (2022) "Imagery, Maps and Monitoring of the World's Tropical Coral Reefs." https://zenodo.org/record/6622015

Álvarez-Alegría, Miguel & Moreno-Spiegelberg, Pablo & Matías, Manuel & Gomila, Damià. (2025). Excitable dynamics and coral reef formation: A simple model of macro-scale structure development. https://doi.org/10.1103/PhysRevResearch.7.023196

Aubert, Olivier. "Origin and stratigraphic evolution of the Maldives (central Indian Ocean)." (1994) Diss., Rice University. https://hdl.handle.net/1911/16703.

Aubert, O., & Droxler, A. W. (1996). Seismic stratigraphy and depositional signatures of the Maldive carbonate system (Indian Ocean). *Marine and Petroleum Geology, 13*(5), 503-536. https://doi.org/10.1016/0264-8172(96)00008-6

Betzler C, Christian Hübscher, Sebastian Lindhorst, John J.G. Reijmer, Miriam Römer, André W. Droxler, Jörn Fürstenau, Thomas Lüdmann; Monsoon-induced partial carbonate platform drowning (Maldives, Indian Ocean). *Geology* 2009;; 37 (10): 867-870. doi: https://doi.org/10.1130/G25702A.1

Betzler C, Eberli GP, Kroon D, Wright JD, Swart PK, Nath BN, Alvarez-Zarikian CA, Alonso-García M, Bialik OM, Blättler CL, Guo JA, Haffen S, Horozal S, Inoue M, Jovane L, Lanci L, Laya JC, Mee AL, Lüdmann T, Nakakuni M, Niino K, Petruny LM, Pratiwi SD, Reijmer JJ, Reolid J, Slagle AL, Sloss CR, Su X, Yao Z, Young JR (2016). The abrupt onset of the modern South Asian Monsoon winds. Sci Rep. 2016 Jul 20;6:29838. doi: 10.1038/srep29838

Bradbury, Roger & Reichelt, Russell. (1983). Fractal Dimension of a Coral Reef at Ecological Scales. Marine Ecology-progress Series - MAR ECOL-PROGR SER. 10. 169-171. 10.3354/meps010169.

Bythell, J.C., Brown, B.E. and Kirkwood, T.B.L. (2018), Do reef corals age?. Biol Rev, 93: 1192-1202. https://doi.org/10.1111/brv.12391

Comeau, S., Edmunds, P. J., Lantz, C. A., & Carpenter, R. C. (2014). *Water flow modulates the response of coral reef communities to ocean acidification*. Scientific Reports, 4, 6681. https://doi.org/10.1038/srep06681

Darwin, Charles. The structure and distribution of coral reefs. Vol. 15. D. Appleton, 1889.

Droxler, A.W. & Jorry, S.J. (2021). The origin of modern atolls: Challenging Darwin's deeply ingrained theory. *Annual Review of Marine Science*, 13, 537-573. doi:10.1146/annurev-marine-122414-034137.

Basillais É. (1997), Coral surfaces and fractal dimensions: a new method, Comptes Rendus de l'Académie des Sciences - Series III - Sciences de la Vie, Volume 320, Issue 8,1997, Pages 653-657, ISSN 0764-4469, https://doi.org/10.1016/S0764-4469(97)85699-5.

Giménez-Romero, À., Matías, M.A. and Duarte, C.M. (2025), *Unravelling the Universal Spatial Properties of Coral Reefs*. Global Ecol Biogeogr, 34: e13939. https://doi.org/10.1111/geb.13939

Graham, N. A. J. (2014). *Habitat complexity: Coral structural loss leads to fisheries declines*. *Current Biology, 24*(9), R359-R361. https://doi.org/10.1016/j.cub.2014.03.069

Grigg, R. W. (1998). Holocene coral reef accretion in Hawaii: a function of wave exposure and sea level history. *Coral Reefs, 17*(3), 263-272. https://doi.org/10.1007/s003380050127

Guo, Z., Zheng, C., & Shi, B. (2002). *Discrete lattice effects on the forcing term in the lattice Boltzmann method*. Physical Review E, 65(4), 046308. https://doi.org/10.1103/PhysRevE.65.046308

Houlbrèque, F., & Ferrier-Pagès, C. (2009). *Heterotrophy in tropical scleractinian corals*. Biological Reviews, 84(1), 1-17. https://doi.org/10.1111/j.1469-185X.2008.00058.x

Jokiel, Paul. (1991). JOKIEL'S ILLUSTRATED SCIENTIFIC GUIDE TO KANEʻOHE BAY, OʻAHU. https://doi.org/10.13140/2.1.3051.9360

Ferrario, F., Beck, M. W., Storlazzi, C. D., Micheli, F., Shepard, C. C., & Airoldi, L. (2014). *The effectiveness of coral reefs for coastal hazard risk reduction and adaptation*. Nature Communications, 5, 3794. https://doi.org/10.1038/ncomms4794 .

van de Koppel, J., Rietkerk, M., Dankers, N., & Herman, P. M. J. (2005). Scale-dependent feedback and regular spatial patterns in young mussel beds. *The American Naturalist, 165*(3), E66-E77. https://doi.org/10.1086/428362

van de Koppel J, Gascoigne JC, Theraulaz G, Rietkerk M, Mooij WM, Herman PM . Experimental evidence for spatial self-organization and its emergent effects in mussel bed ecosystems. Science. 2008 Oct 31;322(5902):739-42. doi: https://doi.org/10.1126/science.1163952

Kroon, D., Steens, T., & Troelstra, S. R. (1991). Onset of monsoonal related upwelling in the western Arabian Sea as revealed by planktonic foraminifers. In W. L. Prell, N. Niitsuma, et al. (Eds.), *Proceedings of the Ocean Drilling Program, Scientific Results* (Vol. 117, pp. 257-263). Ocean Drilling Program. https://doi.org/10.2973/odp.proc.sr.117.126.1991

Krüger, T., Kusumaatmaja, H., Kuzmin, A., Shardt, O., Silva, G., & Viggen, E. M. (2017). *The Lattice Boltzmann Method: Principles and Practice*. Springer. https://doi.org/10.1007/978-3-319-44649-3

Ladd, H. S., Tracey, J. I., & Gross, M. G. (1953). *Drilling on Eniwetok Atoll, Marshall Islands. U.S. Geological Survey Professional Paper 680-A*, 23 p. Washington, D.C.: U.S. Government Printing Office. https://doi.org/10.31e other 33/pp260Y

Ladd, H. S., Tracey, J. I., & Gross, M. G. (1970). Drilling operations on Midway Atoll, Hawaii. *U.S. Geological Survey Professional Paper 680-A*. Washington, D.C.: U.S. Government Printing Office. https://doi.org/10.3133/pp680A

Madin, J.S., Baird, A.H., Dornelas, M., Connolly, S.R., 2014. Mechanical vulnerability explains size-dependent mortality of reef corals. Ecol. Lett. 17, 1008-1015. https:// doi.org/10.1111/ele.12306.

Mandelbrot, Benoit B. *The Fractal Geometry of Nature*. New York: W. H. Freeman and Company, 1983. 3rd ed. ISBN 978-0-7167-1186-5.

Mass, T., Genin, A., Shavit, U., Grinstein, M., Tchernov, D., & Falkowski, P. G. (2010). Flow Enhances Photosynthesis in Marine Benthic Autotrophs by Increasing the Efflux of Oxygen from the Organism to the Water. *Proceedings of the National Academy of Sciences of the United States of America*, *107*(6), 2527-2531. http://www.jstor.org/stable/40536618

Mistr, S. and Bercovici, D. (2003) A Theoretical Model of Pattern Formation in Coral Reefs. Ecosystems, 6, 61-74. http://dx.doi.org/10.1007/s10021-002-0199-0

Mori, T., Smith, T. E., & Hsu, W.-T. 2020. "Common power laws for cities and spatial fractal structures." *Proceedings of the National Academy of Sciences* 117(12): 6469-6475. https://doi.org/10.1073/pnas.1913014117

Naseer, A., & Hatcher, B. G. (2001). Assessing the integrated growth response of coral reefs to monsoon forcing using morphometric analysis of reefs in the Maldives. *In Proceedings of the 9th International Coral Reef Symposium*, Bali, Indonesia, 23-27 October 2000.

Niemeyer, L., Pietronero, L., & Wiesmann, H. J. (1984). *Fractal dimension of dielectric breakdown.* Physical Review Letters, 52(12), 1033-1036. https://doi.org/10.1103/PhysRevLett.52.1033

Rietkerk, M., Boerlijst, M. C., van Langevelde, F., HilleRisLambers, R., de Koppel, J. van, Kumar, L., Prins, H. H. T., de Roos, A. M., & Associate Editor: Mark Westoby. (2002). Self-Organization of Vegetation in Arid Ecosystems. *The American Naturalist*, *160*(4), 524-530. https://doi.org/10.1086/342078

Rietkerk M, van de Koppel J. Regular pattern formation in real ecosystems. Trends Ecol Evol. 2008 Mar;23(3):169-75. doi:10.1016/j.tree.2007.10.013

Rinkevich, B., Loya, Y (1986). Senescence and dying signals in a reef building coral. *Experientia* 42, 320-322. https://doi.org/10.1007/BF01942521

Roy, K.J. 1970. Change in bathymetric configuration, Kaneohe Bay, Oahu, 1882-1969. Hawaii Institute of Geophysics, University of Hawaii, Report 70-15: 26 pp.

Ruiz-Reynés, Daniel & Gomila, Damià & Sintes, Tomàs & Hernández-García, Emilio & Marba, Nuria & Duarte, Carlos. (2017). Fairy circle landscapes under the sea. Science Advances. 3. e1603262. DOI: 10.1126/sciadv.1603262

D. Ruiz-Reynés, E. Mayol, T. Sintes, I.E. Hendriks, E. Hernández-García, C.M. Duarte, N. Marbà, & D. Gomila, Self-organized sulfide-driven traveling pulses shape seagrass meadows, Proc. Natl. Acad. Sci. U.S.A. 120 (3) e2216024120, https://doi.org/10.1073/pnas.2216024120 (2023).

Schlager, W. (2005). *Looking back, moving forward*. In W. Schlager, *Carbonate sedimentology and sequence stratigraphy* (pp. 157-166). SEPM Society for Sedimentary Geology. https://doi.org/10.2110/csp.05.08.0157

Schlager, W. and Purkis, S. (2015), Reticulate reef patterns - antecedent karst versus self-organization. Sedimentology, 62: 501-515. https://doi.org/10.1111/sed.12172

Seekell, D. A., Pace, M. L., Tranvik, L. J., & Verpoorter, C. (2013). "A fractal-based approach to lake size-distributions." *Geophysical Research Letters* 40(3): 517-521. https://doi.org/10.1002/grl.50139

Succi, S. (2001). *The Lattice Boltzmann Equation for Fluid Dynamics and Beyond*. Oxford University Press. ISBN: 978-0-19-850398-9

van de Vijsel, R.C., Hernández-García, E., Orfila, A. *et al.* Optimal wave reflection as a mechanism for seagrass self-organization. *Sci Rep* 13, 20278 (2023). https://doi.org/10.1038/s41598-023-46788-4

Purdy, E. G., & Bertram, G. T. (1993). *Carbonate Concepts from the Maldives, Indian Ocean.* AAPG Studies in Geology, 34. DOI:10.1306/st34568

Purkis, S. J., K. E. Kohler, B. M. Riegl, and S. O. Rohmann. 2007. "The Statistics of Natural Shapes in Modern Coral Reef Landscapes." *Journal of Geology* 115: 493-508. https://doi.org/10.1086/519774

Purkis S, Casini G, Hunt D, Colpaert A. (2015) Morphometric patterns in Modern carbonate platforms can be applied to the ancient rock record: Similarities between Modern Alacranes Reef and Upper Palaeozoic platforms of the Barents Sea, Sedimentary Geology, Volume 321, 2015, Pages 49-69, https://doi.org/10.1016/j.sedgeo.2015.03.001.

Smith, B. T., Frankel, E., & Jell, J. S. (1998). Lagoonal sedimentation and reef development on Heron Reef, southern Great Barrier Reef Province. In G. F. Camoin & P. J. Davies (Eds.), *Reefs and Carbonate Platforms in the Pacific and Indian Oceans* (pp. 279-294). https://doi.org/10.1002/9781444304879.ch15

Sous, D., Bouchette, F., Doerflinger, E., Meulé, S., Certain, R., Toulemonde, G., Dubarbier, B., and Salvat, B. (2020). "On the Small-Scale Fractal Geometrical Structure of a Living Coral Reef Barrier." *Earth Surface Processes and Landforms* 45: 3042-3054. https://doi.org/10.1002/esp.4950

Stoddart, D.R.; Scoffin T.P. (1979). "Microatolls: review of form, origin and terminology" (PDF). *Atoll Research Bulletin*. **224**: 1-17. doi:10.5479/si.00775630.224.1

Xi, H., Dong, X., Chirayath, V. *et al.* Emergent coral reef patterning via spatial self-organization. *Coral Reefs* 44, 273-289 (2025). https://doi.org/10.1007/s00338-024-02603-8

Zawada DG, Brock JC "A Multiscale Analysis of Coral Reef Topographic Complexity Using Lidar-Derived Bathymetry," Journal of Coastal Research, 2009 (10053), 6-15, (1 November 2009)

Zawada, K.J.A., Dornelas, M. & Madin, J.S. Quantifying coral morphology. *Coral Reefs* 38, 1281-1292 (2019). https://doi.org/10.1007/s00338-019-01842-4

Zheng, H., Powell, C. M., Rea, D. K., Wang, J., & Wang, P. (2004). Late Miocene and mid-Pliocene enhancement of the East Asian monsoon as viewed from the land and sea. *Global and Planetary Change, 41*(3-4), 147-155. https://doi.org/10.1016/j.gloplacha.2004.01.003

Zou, Q., & He, X. (1997). On pressure and velocity boundary conditions for the lattice Boltzmann BGK model. *Physics of Fluids*, 9(6), 1591-1598. https://doi.org/10.1063/1.869307

**Acknowledgments**

T.S, A.G.R and E.L. acknowledge financial support from the project acknowledge support by grants PID2021-123723OB-C22 (CYCLE) and PID2024-156062OB-I00 (CHANGE-ME) funded by the Spanish Ministry of Science and Innovation MICIU/AEI/10.13039/501100011033 and by ERDF, EU;  and CEX2021-001164-M (María de Maeztu Program for Units of Excellence in R&D) funded by MICIU/AEI/10.13039/501100011033. E.L. acknowledges funding from the European Union's Horizon 2021 research and innovation programme under the Marie Skłodowska-Curie grant agreement no. 101063295 (CoralMath). A.G.R. acknowledges financial support from grant JDC2024-053275-I, funded by  MICIU/AEI/10.13039/501100011033 and FSE+, and from grant PID2024-156062OB-I00 (CHANGE-ME), funded by the Spanish Ministry of Science and Innovation MICIU/AEI/10.13039/501100011033 and by ERDF, EU. C.M.D. acknowledge the funding from King Abdullah University of Science and Technology (KAUST) baseline fundings BAS/1/1071-01-01.

**Supplementary Materials A. LBM rescaling and hydrodynamic parameterisation**

A major advantage of the lattice Boltzmann method (LBM) is that it is formulated in lattice units and therefore naturally supports systematic rescaling to physical units once a length scale is specified. Table S1 illustrates this procedure by listing the values of the main hydrodynamic parameters. The fourth row shows the values expressed in LBM units, where the lattice spacing and time step are set to unity ($\Delta x_{LBM} = \Delta t_{LBM} = 1$). These quantities are not meant to match physical magnitudes directly, but rather provide a numerically stable set of parameters from which dimensionful values can be obtained. To convert to physical units, we choose the physical lattice spacing $\Delta x_{phys}$ and fix a target physical kinematic viscosity $\nu_{phys}$, which set uniquely determines the physical time step via:

$$\Delta t_{phys} = \Delta t_{LBM} \left( \frac{\Delta x_{phys}}{\Delta x_{LBM}} \right)^2 \frac{\nu_{LBM}}{\nu_{phys}} \tag{1}$$

after which velocities and diffusivities follow from standard scaling rules. Here we consider three lattice spacings (Cases I–III) while holding the physical viscosity fixed at $\nu_{phys} = 1.15 m^2/s$; the corresponding physical parameter values obtained from this mapping are reported in the last three rows of Table S1.

Importantly, this rescaling should be interpreted as a change of units rather than a change of physics. Model behaviour is controlled by the dynamical regime—namely, the relative balance between advection, diffusion, and viscous stresses—rather than the absolute values of individual parameters. This regime is captured by the Reynolds number ($Re = U_{in}L_x/\nu$) and the Péclet number ($Pe = U_{in}L_x/D$), which are dimensionless and remain unchanged across representations. In our simulations, all Case I-III yield the same values, $Re \approx 21.7$ and $Pe \approx 56.3$. These values place the system in a laminar flow regime (low Reynolds number, smooth flow without turbulence) and an advection-dominated transport regime (Péclet number moderately larger than 1, with flow carrying resources more effectively than molecular diffusion). Although these values are lower than those in the open ocean (typically $Re \sim 10^5 - 10^6$ and $Pe \sim 10^2 - 10^3$), they ensure that the dominant physical mechanisms are preserved, allowing the model to capture the essential dynamics of resource transport across reefs focusing on advection and non-turbulent regimes.

Moreover, the specific physical magnitudes reported in Table S1 are of secondary importance within our modelling framework. In the biological update rules (Eqs. 1–3 in the main text), the resource concentration $R(\vec{x})$ and flow speed $|\vec{u}(\vec{x})|$ enter only through normalized probability weights. As a result, it is the relative spatial contrasts in resource availability and hydrodynamic exposure that govern growth and mortality, rather than their absolute value. This further reinforces that our results depend on the hydrodynamic regime and spatial structure of the fields, not on the precise physical calibration of individual parameters.

**Supplementary Materials B. Mathematical background for area–perimeter relations in constant-width shapes**

This appendix summarises the geometric principles underlying the area–perimeter scaling of two simple classes of shapes: annuli and bands (ribbons). In both cases, the shape has a fixed width $w$, while its characteristic extent (radius or length) grows. These idealised geometries provide a baseline for understanding why constant-width structures generally produce a linear relation between area and perimeter in log-scale.

### B.1. Annuli of fixed width

Consider an annulus whose width, $w = R_2 - R_1$, remains constant as the radii grow. Introducing the mean radius, $R = (R_1 + R_2)/2$, the area and perimeter may be written compactly as

$$A = 2\pi R w \qquad P = 4\pi R \tag{2}$$

Both quantities grow linearly with the characteristic size $R$, therefore the area–perimeter scaling is:

$$A = \frac{w}{2} P \tag{3}$$

### B.2. Ribbons of fixed width

A rectangular band of width $w$ and length $L$ satisfies

$$A = Lw \qquad P = 2L + 2w \tag{4}$$

For sufficiently long bands ($L \gg w$), the constant term becomes negligible and the scaling reduces to:

$$A = \frac{w}{2} P \tag{5}$$

### B.3. Log–log representation

Both geometries can be expressed as a power law $A = c P^\alpha$, with $\alpha = 1$ and $c = w/2$. Taking logarithms yields

$$\log A = \log\left(\frac{w}{2}\right) + \log P \tag{6}$$

Thus, the slope of the line in log–log space is $\alpha = 1$ and the intercept is $\log(2/w)$, showing that the width $w$ determines the vertical offset of the line.

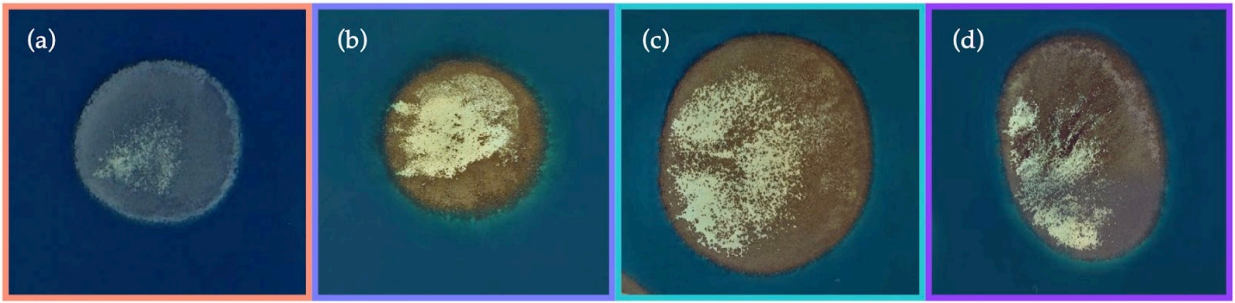

**Figure S1. Patch reefs in Kāneʻohe Bay, Hawaiʻi.** High-resolution satellite imagery illustrating characteristic shapes and directional rim asymmetries. Panels are shown in plan view with north oriented upward. Scale is not standardised across panels, but patch reefs sizes are typically ~100–200m in size. Reefs are broadly circular to oval in outline and often show partially depleted interiors and asymmetric rims, with thicker reef development on the north-east–facing flanks, aligned with the prevailing current direction, and thinner rims on the leeward sides. Images were obtained from Google Earth Pro (acquisition year 2023–2024).

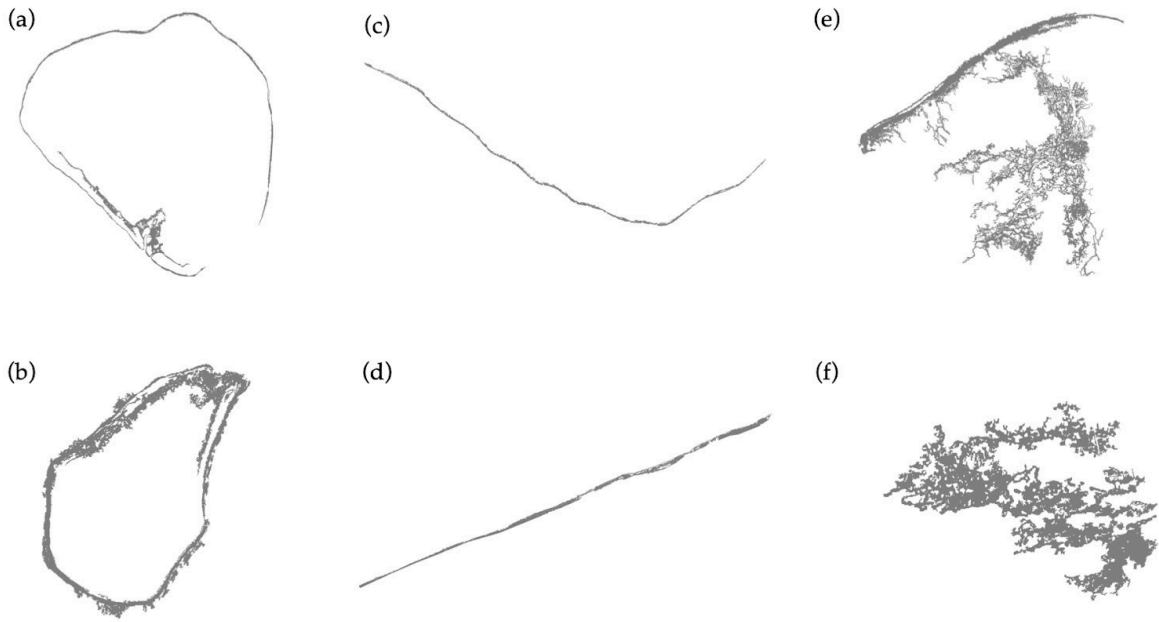

**Figure S2. Representative coral-reef planforms from the Allen Coral Atlas illustrating common large-scale morphologies across regions.** (a–b) Annular, atoll-like reefs; (c–d) linear barrier/fringing reef systems; (e) mixed morphology combining a barrier-like rim with pronounced reticulate structures in the inner lagoon; (f) labyrinthine, maze-like structures within a lagoon. Examples are drawn from Western Micronesia (a), the central Indian Ocean (b), the South Pacific (c), eastern Africa (d), the Hawaiian Islands (e), and Bermuda (f). Panels are not to scale and were selected qualitatively, to provide representative examples of these recurring morphologies.

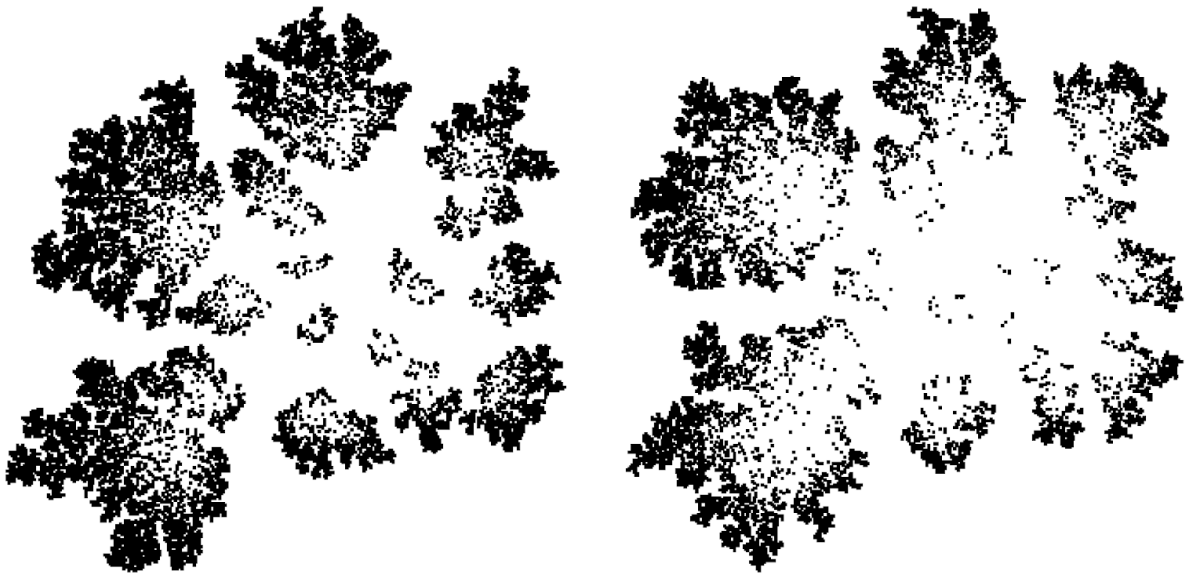

**Figure S3. Dendritic, flow-oriented branching.** Zoomed detail of the simulated reef silhouettes shown in Fig. 5a, highlighting dendritic morphologies. Flow is imposed from left to right. Left: $\omega_{res} = 0.5$. Right: $\omega_{res} = 0.7$. Both panels correspond to the low-erosion regime ($\omega_{ero} = 0$).